# Direct Measurement of the Radiation Pattern of a Nanoantenna Dipole Array

J.Stokes, P.Bassindale, J.W.Munns, Y.Yu, G.S.Hilton, J.R.Pugh, A.Yang, A.Collins, P.J.Heard, R.Oulton, A. Sarua, M.Kuball, A.J.Orr-Ewing and M.J. Cryan
University of Bristol, Bristol, UK.

Z.H.Yuan
Guizhou University For Nationalites,
Guiyang, Guizhou, China

*Abstract*—Scanning microphotoluminescence is used to characterise the fluorescence from a dye-loaded polymer deposited on a 5 x 5 nanoantenna dipole array. Vertical and horizontal scans show anisotropic emission patterns.

*Keywords-component : Surface Plasmon;Nanoantennas*

## I. INTRODUCTION

Optical nanoantennas are revolutionising optics and photonics with their ability to dramatically modify the emission and absorption of light [1,2]. Numerous applications such as sensing, solar energy and quantum information processing can benefit from their unique attributes such as sub-diffraction limit focusing of light and "impedance matching" from quantum emitters to free space. Much work has been done on single antennas, but only recently have antenna arrays begun to be studied in depth. As with their more conventional RF counterparts moving from single elements to an array opens up numerous exciting possibilities such as radiation pattern shaping, much increased directionality[2] and even possibilities for real time beam steering as in a phased array antenna. This paper presents one of the first examples of a measurement of a 2D array of dipole nanoantennas. It uses a well established technique of enhanced emission from a dye-polymer which is spun onto the antenna array[3]. The important contribution here is the use of a high precision scanning microphotoluminescence (MicroPL) set up which enables far field scans across the array to be carried out in both vertical and horizontal directions with ~100nm spatial step size. This enabled the anisotropy of the far field to be observed and will allow a detailed study of numerous array effects such as element spacing and inter-element coupling.

## II. RESULTS

A single element dipole was designed to be resonant in the emission range of LDS 798 dye (www.exciton.com). The 3D Finite Difference Time Domain method was used to design the antenna and Figure 1(a) shows the dependence of the resonant wavelength on arm length for a gold-on-glass dipole antenna with a gap of 30nm, an arm width of 40nm and thickness of 50nm. Drude-Lorentz fitting was used to account for the strong dispersion in the dielectric constant of gold. A FDTD mesh of 5nm was used, which is a good compromise between accuracy and run time. The simulations were performed on University of Bristol's supercomputer : BlueCrystal. Figure 1(a) shows the $E_x$ field 400nm above the centre of the antenna which is excited with a diagonally oriented point dipole placed at the centre of the antenna gap. The dipole alignment relative to the antenna is important in determining the overall enhancement of emission and in the case of the dye there will be an ensemble average effect. The results show that for a range of arm lengths these antennas should strongly enhance emission from a dye emitting in this wavelength range.

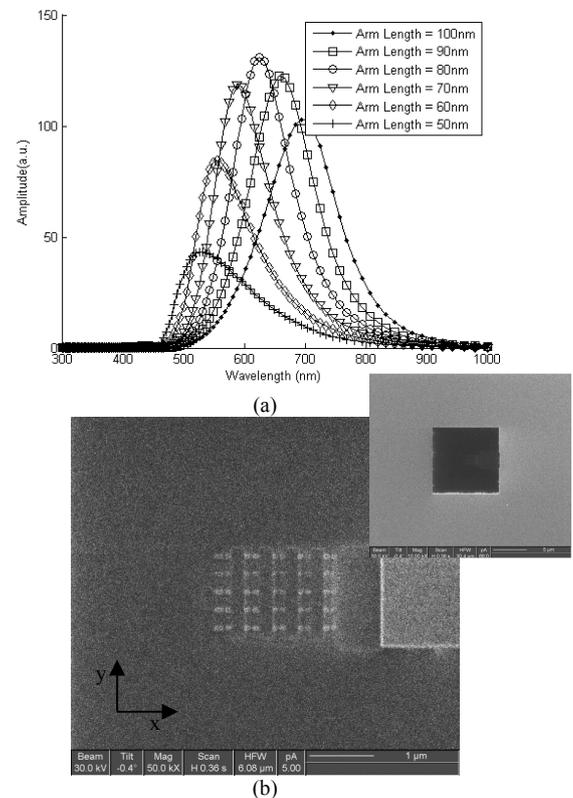

Fig. 1. (a) FDTD modelling of *x*-directed electric field 400nm above the centre of a single dipole antenna for different arm lengths (b) FIB image of a FIB fabricated 5 x 5 dipole array with arm length=100nm, gap=30nm, arm width 40nm, on 50nm thick gold, *x* centre-centre spacing = 400nm, *y* spacing=240nm (Inset) Zoom out of FIB etched region

An antenna array was then fabricated using Focused Ion Beam (FIB) etching. A glass slide coated with 50nm thick gold was used and a procedure was developed to allow a step-by-step definition of the 5x5 array. Care must be taken to reduce charging effects which occur on isolated elements, thus connection to the main gold film was maintained until the very last stage. Figure 1 (b) shows the FIB image of the array along with the side connection arm that has been disconnected from the array and is eventually removed to leave an isolated array within a 10x10μm area where all gold has been removed. The array was then spin coated with dye-loaded polymer. A mixture of LDS 798 in Acetone (0.2 M) and PMMA (in Toluene, 5 vol%) was prepared. 75μl was dropped onto the centre of the antennas and spun for 10sec at 6000rpm (SPS spin150 www.sps-Europe.com) to achieve an even coating. Stabilisation by the polymer led the emission to be red shifted to 680nm. An AFM scan across the region containing the antennas after polymer spinning showed a flatness within 2-3nm, thus preventing any unwanted lensing effects. The dye loaded arrays were then placed in a Renishaw InVia micro-PL system equipped with scanning microscope stage and optical spectrometer with a CCD detector. To excite fluorescence spectra a 532nm diode laser was used and spectra were collected using 100x NA=0.9 objective lens. The laser spot size on the sample was about 0.35 μm and XY scans were performed with 0.1 μm step resolution.

Figure 2 (a) and (b) show vertical and horizontal scans of fluorescence intensity across the array vs wavelength and beam position. The main fluorescence peak from the dye was observed at around 677 nm and the antenna array location was around origin of the line scan. In both cases when the dye was excited over the bulk gold film enhanced emission was observed due to both reflection and surface plasmon effects which can occur when the dye molecules are close to the gold surface. While emission drops in the gap between antennas and bulk gold film there is a pronounced enhancement in emission as the scan reaches the antenna array. The emission from the antenna array appears to be stronger than that for the bulk gold manifesting the enhancement due light-plasmon interaction in the nanoantennas. Detailed calibration runs are underway to confirm this and the exact enhancement ratio. Comparing Figures 2(a) and (b) it can be seen that the beam width is different in each case being approximately ~3μm in the $y$ scan and ~1μm in the $x$ scan. There are two effects occurring here (i) dipole antennas have very anisotropic patterns and it is expected that this will manifest itself in the array radiation pattern (ii) antenna array patterns are strongly dependent on array element spacing, here the $x$ centre to centre spacing is ~400nm whereas the $y$ spacing is ~240nm. This will tend to narrow the beam in the $x$ direction as observed in Figure 2. There is also a slight asymmetry in the scan from one side to the other probably related to a focusing change across large distance or to gold film quality. These scans also show the wavelength response, however, the fairly low quality factor of the antenna enhancement as shown in Figure 1(a) results in very little effect on the dye emission spectrum.

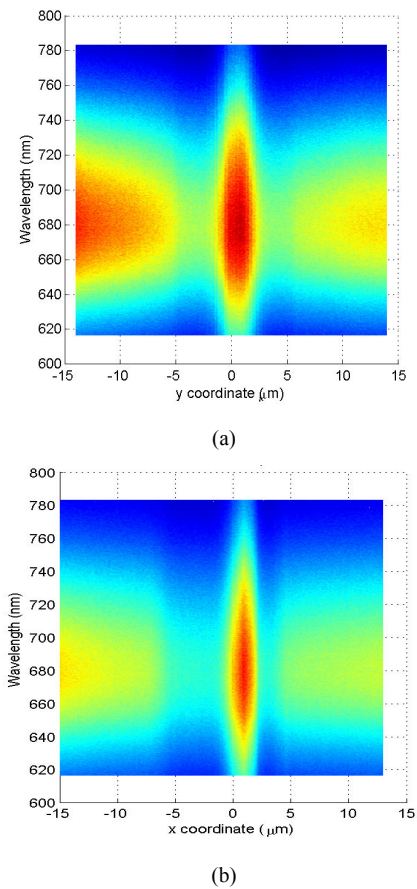

(a)

(b)

Fig. 2. Microphotoluminescence scan, array position is near $y$=0μm, $x$=0μm, FIB etched region is from -5μm to +5μm (a) y-axis (b) x-axis

III. CONCLUSIONS

This paper has presented one of the first direct measurements of the 2D radiation pattern of a dipole nanoantenna array. MicroPL allows an accurate assessment of the far field pattern and full 2D scans are now underway. This approach will allow a very detailed study of array effects such as element spacing and inter-element coupling. FDTD modelling is being carried out to further study the relationship between single element and array patterns and microPL scans at increasing pump power will allow the investigation of possible array lasing effects.